\documentclass[11pt]{article}
\usepackage{verbatim}
\usepackage{amsmath,amssymb}
\makeatletter \@addtoreset{equation}{section} \makeatother

\newcommand{\noi}{\vspace{12pt}\noindent}
\newcommand{\beq}{\begin{equation}}
\newcommand{\eeq}{\end{equation}}
\newcommand{\bea}{\begin{eqnarray}}
\newcommand{\eea}{\end{eqnarray}}

\newcommand{\e}[1]{{(\ref{#1})}}
\newcommand{\eq}[1]{{eq.\ (\ref{#1})}}
\newcommand{\es}[2]{{(\ref{#1}) and (\ref{#2})}}
\newcommand{\eqs}[2]{{eqs.\ (\ref{#1}) and (\ref{#2})}}
\newcommand{\Ref}[1]{{Ref.~\cite{#1}}}

\newcommand{\equi}[1]{\stackrel{{#1}}{=}}

\newcommand{\C}{\mathbb{C}}

\newcommand{\R}{\mathbb{R}}

\newcommand{\ie}{{${ i.e., \ }$}}
\newcommand{\eg}{{${ e.g., \ }$}}

\newcommand{\cf}{{cf.\ }}

\newcommand{\wrt}{{with respect to }}

\renewcommand{\propto}{\sim}
\renewcommand{\~}{ \ }
\renewcommand{\=}{ \ = \ }

\newcommand{\p}{\!{}^{}}
\newcommand{\q}{{}^{}}

\newcommand{\cp}{product }

\newcommand{\for}{{\rm for}}
\newcommand{\hf}{{\scriptstyle{\frac{1}{2}}}}
\newcommand{\Hf}{\frac{1}{2}}

\newcommand{\twostack}[2]{\begin{array}{c} \lower.8ex\hbox{${#1}$}
                     \cr \raise.8ex\hbox{${#2}$} \end{array}}
\newcommand{\deder}[1]{\frac{ 
 \stackrel{\raise.2ex\hbox{$\leftarrow$}}{\delta^{r}}   } 
 {   \delta {#1}}  }
\newcommand{\dedel}[1]{\frac{ 
 \stackrel{\lower.3ex \hbox{$\rightarrow$}}{\delta^{\ell}}   }
 {   \delta {#1}}  }
\newcommand{\papar}[1]{\frac{  
 \stackrel{\raise.2ex\hbox{$\leftarrow$}}{\partial^{r}}   } 
 {   \partial {#1}}  }
\newcommand{\papal}[1]{\frac{ 
 \stackrel{\lower.3ex \hbox{$\rightarrow$}}{\partial^{\ell}}   }
 {   \partial {#1}}  }
\newcommand{\rpa}[1]{{ 
 \stackrel{\raise.2ex\hbox{$\leftarrow$}}{\partial^{r}_{#1}}   }}
\newcommand{\lpa}[1]{{ 
 \stackrel{\lower.3ex\hbox{$\rightarrow$}}{\partial^{\ell}_{#1}}  }}

\newcommand{\proofbox}{\begin{flushright}{\hfill \ensuremath{\Box}}
\end{flushright}}

\newtheorem{theorem}{Theorem}[section]

\newtheorem{lemma}[theorem]{Lemma}

\begin{document}
\thispagestyle{empty}
\title{\Large{\bf Semiclassical Double-Inequality \\ 
on Heisenberg Uncertainty Relation \\ in 1D}}
\author{{\sc Klaus~Bering}$^{a}$ \\~\\
Institute for Theoretical Physics \& Astrophysics\\
Masaryk University\\Kotl\'a\v{r}sk\'a 2\\CZ--611 37 Brno\\Czech Republic}
\maketitle
\vfill
\begin{abstract}
We prove a double-inequality for the product of uncertainties for position 
and momentum of bound states for 1D quantum mechanical systems in the 
semiclassical limit. 
\end{abstract}
\vfill
\begin{quote}
PACS number(s): 03.65.-w; 03.65.Ge; 03.65.Sq; 
\hrule width 5.cm \vskip 2.mm \noindent 
$^{a}${\small E--mail:~{\tt bering@physics.muni.cz}} \\
\end{quote}

\newpage

\section{Introduction}
\label{secintro}

\noi
It is known that the \cp of position and momentum uncertainties for the 
$N$'th bound states of (i) the harmonic oscillator and (ii) the infinite 
square well is exactly given as
\beq 
\frac{\Delta x\~\Delta p}{\hbar} \= N+\Hf\~, \qquad N\~\in\~\mathbb{N}\q_{0}\~,
\label{sho01}
\eeq
and
\beq 
\frac{\Delta x\~\Delta p}{\hbar} \= \Hf\sqrt{\frac{(\pi N)^{2}}{3}-2}\~, 
\qquad N\~\in\~\mathbb{N}\~,  \label{inftywell01}
\eeq
respectively. In particular, they display a linear dependence of $N$ for 
$N \gg 1$. The asymptotic slope $\frac{\pi}{2\sqrt{3}}\approx 0.9069$ of 
the infinite square well \e{inftywell01} is less than $10\%$ smaller than the
corresponding slope ($=1$) of the harmonic oscillator \e{sho01}. 
It is natural to ponder if there (for quantum mechanical systems in 1D) 
exists a semiclassical double-inequality of the form
\beq 
C\q_{\min} \~\leq\~ U~:=~\frac{\Delta x\~\Delta p}{\hbar N} \~\leq\~C\q_{\max}
\quad\for\quad N\~\gg\~1\~, \label{doublein01}
\eeq 
where $C\q_{\max}> C\q_{\min}>0$ are two dimensionless constants, say, of order 
one? (The upper bound $C\q_{\max}$ cannot be much smaller than one in order not
to conflict with the theoretical Heisenberg uncertainty bound 
$\Delta x\~\Delta p \geq \frac{\hbar}{2}$.)
{}Further physical motivation for such conjecture \e{doublein01} is 
loosely based on the fact that there semiclassically is one bound state per 
phase space area times Planck's constant $\hbar$ \cite{ll3,gp2,friedrich98}. 
See also Gromov's symplectic non-squeezing theorem
\cite{gromov95,degossonluef09}. The \cp of uncertainties in various
examples is also discussed in, \eg \Ref{ushadevikarthik12}.

\noi
Here we are assuming that the system has a large number of bound states, so
that we can apply semiclassical methods.
[On top of the bound states, the system could have a continuum of 
non-normalizable states, which we are not pursuing here. In this article, we 
are only interested in the bound states below the continuum limit $E\q_{0}$. 
Note that $E\q_{0}$ could be $+\infty$.]

\noi
The conjecture in its basic form \e{doublein01} turns out to be false for
at least three reasons (which however may be fixed):

\begin{enumerate}

\item
{}Firstly, it is easy to violate any upper bound $C\q_{\max}$ with a double-well
potential with the two wells separated sufficiently far apart. The remedy is 
to avoid quantum mechanical tunneling, \ie to impose that the classically 
accessible region should be connected, \cf \eq{conninvbranch01}. With this 
assumption (along with some minor technical assumptions, \cf
Section~\ref{secassump}), we shall show that an upper bound is $C\q_{\max}=1$, 
\cf Theorem~\ref{thmupper}. Incidentally, this upper bound is saturated for 
the harmonic oscillator \e{sho01}, \cf \eq{extremell01}.

\item
Secondly, it is possible to violate any non-zero lower bound $C\q_{\min}$ with 
an attractive negative power law potential of the form 
$\Phi(x)\propto |x|^{\epsilon-2}$, where $\epsilon>0$ is an arbitrary small 
number, \cf Appendix~\ref{secnegpower}. The reason is that the spectrum of 
the Hamiltonian is not bounded from below for $\epsilon<0$. Thus close to the
unitarity limit $\epsilon\to 0^{+}$, it is possible to pack arbitrarily
many bound states down the potential throat and saturate the theoretical
Heisenberg uncertainty bound $\Delta x\~\Delta p \geq \frac{\hbar}{2}$.
The remedy is to assume that the potential is bounded from below 
$\Phi(x)\geq V\p_{0}> -\infty$. 

\item
Thirdly, even for a potential that is bounded from below
$\Phi(x)\geq V\p_{0}> -\infty$, any non-zero lower bound $C\q_{\min}$ may be
violated at finite $N\gg 1$, \cf \eg the two-stage infinite well discussed in 
Appendix~\ref{secshallowpot}. The remedy is to consider the infinite 
$N\to\infty$ limit. With these assumptions, we shall show that an lower bound
is $C\q_{\min}=\frac{\pi}{2\sqrt{3}}\approx 0.9069$, \cf Theorem~\ref{thmlower}. 
Incidentally, this lower bound is saturated for the infinite square well
\e{inftywell01}.
\end{enumerate}

\section{Introduction to WKB}
\label{secintrowkb}

\noi
Consider a 1D system with a Hamiltonian of the form
\beq
H(x,p) \= \frac{p^{2}}{2m} + \Phi(x) \~, \qquad x,p\~\in\~\R\~, \label{ham01}
\eeq
where $\Phi:\mathbb{R}\to\mathbb{R}$ denotes the potential energy function. 
For the $N$'th bound state, where the label $N\gg 1$ is large, we can use 
semiclassical WKB approximation methods, \cf \Ref{ll3,gp2,friedrich98}. 
Semiclassically, the number of states $N=N(E)$ below the energy-level $E$ is
given by the area of phase space that is classically accessible, divided by 
Planck's constant $h$,
\beq
 N(E) \~\approx\~ \iint_{H(x,p)\leq E} \frac{dx\~dp}{h}
\=\frac{2}{h}\int_{\Phi(x)\leq E}\! |p(x)| \~dx\~, \label{enn01}
\eeq
where
\beq 
|p(x)|\~:=\~ \sqrt{2m(E-\Phi(x))}\~\geq~0\~.\label{pee01}
\eeq
Since we are only interested in the semiclassical regime, we ignored in 
\eq{enn01} the Maslov index, also known as the metaplectic correction.
(The $\approx$ signs are here to remind us of the semi-classical approximation 
that we made.)
The time-independent Schr\"{o}dinger equation (TISE) is invariant under 
complex conjugation, so we may assume that the bound state wave functions 
are real. The WKB wave function $\psi(x)$ for the $N$'th bound state with 
energy $E$ reads 
\beq
 \psi(x) \~\approx \~ \frac{C}{\sqrt{|p(x)|}}
\cos\left[\frac{S(x)}{\hbar}+\theta \right]\~, 
\label{psi01}
\eeq 
where
\beq 
S(x)\~:=\~\int^{x}_{0} \! dx^{\prime}\~|p(x^{\prime})|\~, \qquad C\~\in\~\C\~,
\qquad \theta~\in\~\R\~.\label{ess01}
\eeq
{}For further justification of the WKB method, we refer to \Ref{gp2}.

\section{Classically Accessible Length}
\label{seccal}

\noi
Let 
\beq 
V\p_{0}\~:=\~ \inf_{x\in\mathbb{R}} \~\Phi(x) \label{veezero01}
\eeq 
be the infimum of the potential energy. ($V\p_{0}$ could be $-\infty$.) Let 
\beq
\ell(V)\~:=\~\lambda(\{x\in\mathbb{R} \mid \Phi(x) \leq V\}) \label{elldef00}
\eeq
be the length of the classically accessible position region at potential 
energy-level $V$. Technically, the length $\ell(V)$ is the Lebesgue 
measure $m$ of the preimage
\beq 
\Phi^{-1}(]-\infty,V])\~:=\~ \{x\in\mathbb{R} \mid \Phi(x) \leq V\}\~,
\label{preimage01}
\eeq
which in principle does not necessarily have to be a connected interval, 
although we will later make this assumption, \cf Section~\ref{secassump}. 
The accessible length must grow with increasing potential energy
\beq
\frac{d\ell(V)}{dV}\~\equiv\~\ell^{\prime}(V)\~\geq\~0\~. \label{ellpos01}
\eeq
The lower potential energy limit 
\beq
V\p_{0}\= \lim_{2x\to0^{+}} \ell^{-1}(\{2x\}) \label{veezero02}
\eeq
satisfies 
\beq
\ell(V\p_{0})\=0\~. \label{bcv0}
\eeq
The continuum limit is 
\beq 
E\q_{0}\~:=\~ \lim_{2x\to\infty} \ell^{-1}(\{2x\})\~. \label{eezero01}
\eeq
We are interested in energies $E\in [V\p_{0},E\q_{0}]$. The immaterial factor 
$2$ that appears in \eqs{veezero02}{eezero01} is spurred by an
assumption \e{conninvbranch01}, which is made later in Section~\ref{secassump}.

\noi
The accumulated accessible length ${\cal L}(V)$ at potential energy-level $V$
is defined as
\beq
{\cal L}(V)\~:=\~\int_{V\p_0}^{V} \ell(V^{\prime})~dV^{\prime}\~.\label{accumell01} 
\eeq

\begin{theorem}[Abel-like integral transform between $N(E)\sim I(E)$ and 
$\ell(V)$] 
The number $N(E)$ of bound states with energy $\leq E$ can be reconstructed 
from the accessible length $\ell(V)$ via the formula
\beq
N(E) \~\approx \~\frac{\sqrt{2m}}{h}I(E)
\~\equiv\~\frac{1}{\pi\hbar}\sqrt{\frac{m}{2}}I(E)\~, \label{enn02}
\eeq
where $I(E)$ is an integral
\beq
 I(E)\~:=\~ 2\int_{\Phi(x)\leq E}\! \sqrt{E-\Phi(x)} \~dx
\=2\int_{V\p_{0}}^{E} \sqrt{E-V}\~\ell^{\prime}(V)\~dV
\~\equi{\e{bcv0}}\~\int_{V\p_{0}}^{E} \frac{\ell(V)\~dV}{\sqrt{E-V}}\~. 
\label{eye01}
\eeq
Conversely, the accumulated accessible length ${\cal L}(V)$ at potential
energy level $V$ can be reconstructed from $I(E)$ via the formula
\beq
{\cal L}(V) 
\=\frac{1}{\pi}\int_{V\p_{0}}^{V} \frac{I(E)\~dE}{\sqrt{V-E}}
\=\frac{2}{\pi}\int_{V\p_{0}}^{V} \!dE\~I^{\prime}(E)\sqrt{V-E}\~.
\label{accumell02}
\eeq
By differentiation of \eq{accumell02}, the accessible length $\ell(V)$ at 
potential energy level $V$ can be reconstructed from $I(E)$ via the formula
\beq 
\ell(V)\~\equiv\~ \frac{d{\cal L}(V)}{dV}
\= \frac{1}{\pi}\frac{d}{dV}\int_{V\p_{0}}^{V}\frac{I(E)\~dE}{\sqrt{V-E}}
\=\frac{1}{\sqrt{\pi}} (D^{\hf}I)(V)
\= \frac{1}{\pi} \int_{V\p_{0}}^{V} \frac{I^{\prime}(E)~dE}{\sqrt{V-E}}
\~. \label{ell02}
\eeq
Here $D^{\hf}$ denotes a fractional derivative, 
$\left(D^{\hf}\right)^{2}=D\equiv \frac{d}{dV}$.
\end{theorem} 

\noi
{\sc Proof of \eq{enn02}}: 
\beq 
h \~N(E) \~\stackrel{\e{enn01}}{\approx}\~ 
2\int_{0}^{\sqrt{2m(E-V\p_{0})}} \ell\left(E-\frac{p^{2}}{2m}\right)\~d|p|
\eeq 
\beq
\~\stackrel{V=E-\frac{p^{2}}{2m}}{=}\~
2\int_{V\p_{0}}^{E} \frac{\ell(V)~dV}{v}
\=\sqrt{2m}\int_{V\p_{0}}^{E}\frac{\ell(V)\~dV}{\sqrt{E-V}}
\~\equi{\e{eye01}}\~\sqrt{2m}I(E)\~,
\eeq
because $dV~=~ - v\~d|p|$ with speed
$v~:=~\frac{|p|}{m}~=~\sqrt{\frac{2(E-V)}{m}}$.
\proofbox

\noi
{\sc Proof of \eq{accumell02}}:~~Notice that 
\beq 
\int_{V^{\prime}}^{V} \frac{dE}{\sqrt{(V-E)(E-V^{\prime})}} 
\~\stackrel{E=V \sin^2\theta + V^{\prime} \cos^2\theta }{=}\~ 
2 \int_{0}^{\frac{\pi}{2}} d\theta \= \pi\~.\label{piint01}
\eeq
Then 
\bea
\int_{V\p_{0}}^{V} \frac{I(E)\~dE}{\sqrt{V-E}} 
&\equi{\e{eye01}}& 
\int_{V\p_{0}}^{V}\frac{dE}{\sqrt{V-E}}\int_{V\p_{0}}^{E}
\frac{\ell(V^{\prime})~dV^{\prime}}{\sqrt{E-V^{\prime}}} \cr
\~\stackrel{{\rm Tonelli}}{=}\~\int_{V\p_0}^V \ell(V^{\prime})\~dV^{\prime}
\int_{V^{\prime}}^{V} \frac{dE}{\sqrt{(V-E)(E-V^{\prime})}}
&\equi{\e{piint01}}&\pi \int_{V\p_0}^V \ell(V^{\prime})~dV^{\prime}
\~\equi{\e{accumell01}}\~\pi {\cal L}(V) \~, 
\label{accumell02proof}
\eea
where we rely on Tonelli's theorem to change the order of integrations.
\proofbox

\section{Momentum averages}
\label{secmom}

\noi
We will use the notation $\langle F \rangle$ to denote the expectation
value of some observable $F$ in the $N$'th bound state. The momentum 
average  
\beq
 \langle p \rangle \=0 \label{peeave1}
\eeq
is automatically zero. The momentum square average becomes
\beq
\langle p^{2} \rangle
\= \int_{\Phi(x)\leq E}\! \left|\hbar \psi^{\prime}(x)\right|^{2}\~dx
\~\stackrel{\e{psi01}}{\approx}\~|C|^{2} \int_{\Phi(x)\leq E}\! |p(x)| 
\sin^{2}\left[\frac{S(x)}{\hbar}+\theta \right]\~dx\~\approx\~ 
\frac{|C|^{2}}{2} \int_{\Phi(x)\leq E}\! |p(x)|\~dx 
\label{peeave2a}
\eeq
in the semiclassical limit $|S(x)| \gg \hbar$. Therefore 
\beq
 \frac{2\langle p^{2} \rangle}{|C|^{2}} 
\~\stackrel{\e{peeave2a}}{\approx}\~\int_{\Phi(x)\leq E}\! |p(x)| \~dx
\~\equi{\e{pee01}+\e{eye01}}\~\sqrt{\frac{m}{2}} I
\~\stackrel{\e{enn02}}{\approx}\~\frac{h}{2}N\~. \label{peeave2b}
\eeq
Similarly, the normalization of the wave function $\psi$ yields
\beq
\frac{2}{|C|^2} 
\~\stackrel{\e{psi01}}{\approx}\~\int_{\Phi(x)\leq E}\!\frac{dx}{|p(x)|}
\~\equi{\e{pee01}+\e{jay01}}\~\frac{J}{\sqrt{2m}}
\~\stackrel{\e{enn02}}{\approx}\~\frac{h}{2m}\frac{dN}{dE}\~,
\label{ceeave2b}
\eeq
where $J(E)$ is an integral
\beq
 J(E)\~:=\~\int_{\Phi(x)\leq E}\! \frac{dx}{\sqrt{E-\Phi(x)}}\=
\int_{V\p_{0}}^{E} \frac{\ell^{\prime}(V)\~dV}{\sqrt{E-V}}
\=I^{\prime}(E) \~.\label{jay01} 
\eeq

\section{Assumptions}
\label{secassump}

\noi
At this stage, to ease calculations, we will from now on make two 
simplifying assumptions:

\begin{enumerate}
\item 
The potential $\Phi$ is an {\bf even} function
\beq 
\Phi(x)\=\Phi(-x)\~.\label{even01}
\eeq
Then the position average
\beq 
 \langle x \rangle \=0 \label{exave1}
\eeq
is zero.

\item 
For all potential energy levels $V$, the classically accessible region is 
{\bf connected}, \ie an interval. 
Then ${\rm sgn}(\Phi^{\prime}(x))={\rm sgn}(x)$, and the accessible length 
\e{elldef00} becomes
\beq
 \ell(V)\=2\Phi^{-1}(V)\label{conninvbranch01}
\eeq
twice the positive inverse branch of $\Phi$. Moreover, the continuum limit 
\e{eezero01} becomes simply
\beq 
E\q_{0}\= \sup_{x\in\mathbb{R}} \~\Phi(x)\~. \label{eezero02}
\eeq 
\end{enumerate}

\noi
Then the formulas for the uncertainties reduce to
\beq
(\Delta x)^{2} \~\equi{\e{exave1}}\~\langle x^{2} \rangle
\quad{\rm and}\quad
(\Delta p)^{2} \~\equi{\e{peeave1}}\~\langle p^{2} \rangle\~.\label{deltaxp01}
\eeq
The position square average becomes  
\beq
 \frac{2\langle x^{2} \rangle}{|C|^{2}} 
\~\stackrel{\e{psi01}}{\approx}\~\int_{\Phi(x)\leq E}\!\frac{x^{2}\~dx}{|p(x)|}
\~\equi{\e{pee01}+\e{kay01}}\~\frac{K}{4\sqrt{2m}}\~,\label{exave2}
\eeq
where $K(E)$ is an integral
\beq
 K(E)\~:=\~\int_{\Phi(x)\leq E}\!\frac{(2x)^{2}\~dx}{\sqrt{E-\Phi(x)}}
\=\int_{V\p_{0}}^{E} \frac{\ell(V)^{2}\ell^{\prime}(V)\~dV}{\sqrt{E-V}} 
\=\int_{V\p_{0}}^{E} \frac{dV}{3\sqrt{E-V}}\frac{d\ell(V)^{3}}{dV}\~. \label{kay01} 
\eeq 
The second equality in \eq{kay01} uses assumption 1 and, in particular, 
assumption 2. Then the \cp of uncertainties reads
\beq
 U\~:=\~\frac{\Delta x\~\Delta p}{\hbar N}
~\approx~ \frac{\pi}{\sqrt{2}J}\sqrt{\frac{K}{I}} \~,\label{you01} 
\eeq
where we used eqs.\ \e{peeave2b}, \e{ceeave2b}, \es{deltaxp01}{exave2}. Note
that the \cp \e{you01} of uncertainties only depends on the three integrals
$I$, $J$, and $K$, which are defined in eqs.\ \e{eye01}, \es{jay01}{kay01},
respectively.

\section{Main Theorems}
\label{secthm}

\noi
We are now ready to state the two main theorems.

\begin{theorem}
[Upper bound] Given assumptions 1 and 2, then the \cp \e{you01} of
uncertainties satisfy the following inequality for large $N\gg 1$:
\beq
  U\~\lesssim\~1\quad\for\quad N\~\gg\~1\~. \label{youthm01} 
\eeq
\label{thmupper}
\end{theorem}

\begin{theorem}
[Lower bound] Given assumptions 1 and 2, 
and if the potential is bounded from below $\Phi(x)\geq V\p_{0}> -\infty$, 
then the \cp \e{you01} of uncertainties satisfy the following inequality in the 
infinite $N\to\infty$ limit:
\beq
 U \~\gtrsim\~\frac{\pi}{2\sqrt{3}}\~\approx\~ 0.9069
\quad\for\quad N\~\to~\infty\~.\label{youthm02} 
\eeq
\label{thmlower}
\end{theorem}

\noi
We stress that the upper bound \e{youthm01} holds for finite $N\gg 1$, while 
this is not necessarily the case for the lower bound \e{youthm02}. See 
Appendix~\ref{sectwostage} for a counterexample.

\noi
We believe that the qualitative picture remains the same if we remove 
assumptions 1, and to some extend, assumption 2.

\section{Extremal profile}
\label{secextremalprofile}

\noi
Note that the independent variable is the derivative $\ell^{\prime}(V)$ rather
than $\ell(V)$ due to the inequality \e{ellpos01}. The first variations read
\beq
\delta I\~\equi{\e{eye01}}\~ 
\int_{V\p_{0}}^{E} \frac{\delta\ell(V)\~dV}{\sqrt{E-V}}
\=2\int_{V\p_{0}}^{E} \sqrt{E-V}\~\delta\ell^{\prime}(V)\~dV 
\=-2\int_{V\p_{0}}^{E} \frac{dV}{\sqrt{E-V}}\frac{d}{dV} 
\left[ (E-V) \delta\ell(V)\right]\~, \label{deltai01}
\eeq
\beq 
\delta J\~\equi{\e{jay01}}\~
\int_{V\p_{0}}^{E} \frac{\delta\ell^{\prime}(V)\~dV}{\sqrt{E-V}}
\~\equi{\e{ubc02}}\~  
-\int_{V\p_0}^{E} \frac{\delta\ell(V)\~dV}{2(E-V)^{\frac{3}{2}}}\~, 
\label{deltaj01}
\eeq
\beq 
\delta K\~\equi{\e{kay01}}\~
\int_{V\p_{0}}^{E} \frac{dV}{\sqrt{E-V}}\frac{d}{dV} 
\left[ \ell(V)^2 \delta\ell(V)\right]
\~\equi{\e{ubc02}}\~  
-\int_{V\p_{0}}^{E} \frac{\ell(V)^2\delta\ell(V)~dV}{2(E-V)^{\frac{3}{2}}}\~.
\label{deltak01}
\eeq
The second variations read 
\beq
\delta^{2}I\~\equi{\e{eye01}}\~0\~\equi{\e{jay01}}\~\delta^{2}J
\label{deltaij02}
\eeq 
(since $I$ and $J$ are linear in $\ell$), and 
\beq
 \delta^{2}K \= 2 \int_{V\p_{0}}^{E} \frac{dV}{\sqrt{E-V}}\frac{d}{dV} 
\left[ \ell(V) \~\delta\ell(V)^{2}\right]
\~\equi{\e{ubc02}}\~ 
-\int_{V\p_{0}}^{E} \frac{\ell(V)\~\delta\ell(V)^2\~dV}{(E-V)^{\frac{3}{2}}}
\~\leq\~0\~. \label{deltak02}
\eeq
[Note that the rewritings of eqs.\ \e{deltaj01}--\e{deltak02} in terms of
$\frac{\delta\ell(V)}{(E-V)^{\frac{3}{2}}}$ are only integrable/meaningful at
the upper limit $V=E$ if we assume the boundary condition
\beq
\delta\ell(V=E)\=0\~,\label{ubc02}
\eeq
which we usually won't assume.] Eqs.\ \e{deltai01}--\e{deltak01} yield the 
first variation
\bea
 \frac{\delta U}{U}
&\equi{\e{you01}}&
\frac{\delta K}{2K} - \frac{\delta I}{2I} - \frac{\delta J}{J}
\=\int_{V\p_{0}}^{E} \frac{dV}{\sqrt{E-V}}\frac{d}{dV} 
\left[ g(V) \delta\ell(V)\right] \cr
&=&\int_{V\p_{0}}^{E} \frac{g(V)\delta\ell^{\prime}(V)\~dV}{\sqrt{E-V}}
+\int_{V\p_0}^{E} \frac{g^{\prime}(V)\~ dV}{\sqrt{E-V}} 
\int_{V\p_{0}}^{V} \!dV^{\prime}\~\delta\ell^{\prime}(V^{\prime}) \cr
&=&\int_{V\p_{0}}^{E} \!dV\~\delta\ell^{\prime}(V) \left[\frac{g(V)}{\sqrt{E-V}}
+\int_{V}^{E} \frac{g^{\prime}(V^{\prime})\~dV^{\prime}}{\sqrt{E-V^{\prime}}}\right]\~,
\label{deltau01}
\eea
where we have defined
\beq
g(V)\~:=\~\frac{\ell(V)^2}{2K}+\frac{E-V}{I}-\frac{1}{J}\~. \label{gee01}
\eeq 
{}From \eq{deltau01} with $\ell^{\prime}(V)$ as independent variable in the
variation, we conclude that the Euler-Lagrange equation reads
\beq
 \frac{g(V)}{\sqrt{E-V}}
+ \int_{V}^{E} \frac{g^{\prime}(V^{\prime}) \~dV^{\prime}}{\sqrt{E-V^{\prime}}}\=0\~. 
\label{eleq01}
\eeq 
Differentiation of \eq{eleq01} \wrt $V$ yields
\beq
 \frac{d}{dV}\left[\frac{g(V)}{\sqrt{E-V}}\right]
\=\frac{g^{\prime}(V)}{\sqrt{E-V}}\~,\label{eleq02}
\eeq
which in turn leads to that an extremal profile satisfies
\beq
 g(V)\=0\~. \label{eleq03}
\eeq
Recalling the definition \e{gee01}, the square $\ell\q_{\ast}(V)^2$ of the
extremal profile must be affine in $V$. (Here the subscript ``$\ast$'' 
denotes extremality.) Together with the boundary condition \e{bcv0} this then
implies that the extremal profile is
\beq
 \ell\q_{\ast}(V)\=A\sqrt{V-V\p_{0}}\~, \qquad A\~>\~0\~,\label{extremell01}
\eeq
which corresponds to a harmonic oscillator $\Phi\q_{\ast}(x)-V\p_{0}
\=\left(\frac{2x}{A}\right)^{2}~\propto~ x^{2}$, \ie a quadratic potential.
The extremal value for the three pertinent integrals are
\beq
 I\q_{\ast}\~\equi{\e{eye01}+\e{extremell01}}\~
A\int_{V\p_{0}}^{E} \frac{\sqrt{V-V\p_{0}}\~dV}{\sqrt{E-V}}
\~\equi{\e{beta01}}\~A(E-V\p_{0})\~B(\frac{3}{2},\frac{1}{2}) 
\= \frac{\pi}{2}A(E-V\p_{0})\~, \label{extremi01}
\eeq
\beq
J\q_{\ast}\~\equi{\e{jay01}+\e{extremell01}}\~
\frac{A}{2}\int_{V\p_{0}}^{E} \frac{dV}{\sqrt{V-V\p_{0}}\sqrt{E-V}}
\~\equi{\e{beta01}}\~
\frac{A}{2} \~B(\frac{1}{2},\frac{1}{2}) 
\= \frac{\pi}{2}A\~,  \label{extremj01}
\eeq
\beq
 K\q_{\ast}\~\equi{\e{kay01}+\e{extremell01}}\~
\frac{A^{3}}{2}\int_{V\p_{0}}^{E} \frac{\sqrt{V-V\p_{0}}\~dV}{\sqrt{E-V}}
\~\equi{\e{beta01}}\~
\frac{A^{3}}{2}(E-V\p_{0})\~B(\frac{3}{2},\frac{1}{2}) 
 \=\frac{\pi}{4}A^{3}(E-V\p_{0})\~,  \label{extremk01}
\eeq
by substitution $v \mapsto V=(E-V\p_{0})v+V\p_{0}$. Here 
\beq
B(x,y)\~=\~\int_{0}^{1}\! dv\~v^{x-1}(1-v)^{y-1} 
\=\frac{\Gamma(x)\Gamma(y)}{\Gamma(x+y)}\~, \qquad 
{\rm Re}(x), {\rm Re}(y)\~>\~ 0\~, 
\label{beta01}
\eeq
is the Euler Beta function. The extremal profile \e{extremell01}
saturates the inequality of the Upper Bound Theorem~\ref{thmupper}
\beq
 U\p_{\ast}\~\equi{\e{you01}+\e{extremell01}}1\~.\label{extremu01}
\eeq

\section{Proof of the Upper Bound Theorem~\ref{thmupper}}
\label{secproofthmupper}

\noi
To prove the Upper Bound Theorem~\ref{thmupper}, we need to check that the 
Hessian is negative semidefinite. At the stationary point, we have
\beq
0\=\frac{\delta U}{U\p_{\ast}}
\~\equi{\e{you01}}\~\frac{\delta K}{2K\q_{\ast}} - \frac{\delta I}{2 I\q_{\ast}}
 - \frac{\delta J}{ J\q_{\ast}}\~, \label{extremdeltau01}
\eeq
or equivalently
\beq
 \frac{\delta K}{K\q_{\ast}}
\= \frac{\delta I}{ I\q_{\ast}} +2 \frac{\delta J}{ J\q_{\ast}}\~. 
\label{extremdeltak01}
\eeq
Using the Cauchy-Schwarz inequality, we derive
\bea
( \delta I)^{2}
&\equi{\e{deltai01}}&
\left[\int_{V\p_{0}}^{E} \frac{\delta\ell(V)\~dV}{\sqrt{E-V}}\right]^{2}
\~\stackrel{\rm CS-ineq.}{\leq}\~
\int_{V\p_{0}}^{E}\frac{\sqrt{E-V}\~dV}{\sqrt{V-V\p_{0}}}
\int_{V\p_{0}}^{E} \frac{\sqrt{V-V\p_{0}}\~\delta\ell(V)^{2}\~dV}{(E-V)^{\frac{3}{2}}} 
\cr
&\equi{\e{deltak02}+\e{extremell01}}&
-\frac{\pi}{2}\frac{E-V\p_{0}}{A}\delta^2 K\~, 
\label{csineq01}
\eea
or equivalently
\beq
\left( \frac{\delta I}{ I\q_{\ast}} \right)^{2} 
\~\stackrel{\e{extremi01}+\e{extremk01}+\e{csineq01}}{\leq}\~ 
-\frac{\delta^{2} K}{2K\q_{\ast}} \~ . \label{niceineq01}
\eeq
Therefore the second variation becomes
\bea
 \frac{\delta^{2}U}{U\p_{\ast}}
&\equi{\e{deltau01}}&\left(\frac{\delta U}{U\p_{\ast}}\right)^{2}
+\frac{\delta^{2} K}{2K\q_{\ast}} 
-\frac{1}{2}\left(\frac{\delta K}{K\q_{\ast}}\right)^{2} 
+\frac{1}{2}\left(\frac{\delta I}{I\q_{\ast}}\right)^{2}
+\left(\frac{\delta J}{J\q_{\ast}}\right)^{2} \cr
&\equi{\e{extremdeltau01}}&\frac{\delta^{2} K}{2K\q_{\ast}} 
-\left(2\frac{\delta I}{I\q_{\ast}} 
+\frac{\delta J}{ J\q_{\ast}}\right) \frac{\delta J}{ J\q_{\ast}}
\~\stackrel{\e{niceineq01}}{\leq}\~
-\left(\frac{\delta I}{I\q_{\ast}} +\frac{\delta J}{ J\q_{\ast}}\right)^{2}
\~\leq\~0\~.\label{extremhessian01}
\eea
Moreover, one may show that the only two zero-modes of the Hessian 
correspond to the two parameters $A$ and $V\p_{0}$ of the harmonic 
potential \e{extremell01}. We conclude that the harmonic potentials 
\e{extremell01} as the only profiles yield the global maximum for $U$.

\section{Hard Wall Potentials}
\label{sechardwall}

\noi
A {\em hard wall potential} is by definition a potential $\Phi$ where the 
classically accessible length $\ell$ is bounded, \ie 
$\exists L<\infty\~\forall V>V\p_{0}:\~ \ell(V) \leq L$. 

\begin{lemma}[Hard Wall Potentials]
If the classically accessible length is bounded and the potential is bounded 
from below $\Phi(x)\geq V\p_{0}>-\infty$, then
\beq
\lim_{E\to\infty}U(E)\~\equi{\e{you01}}\~\frac{\pi}{2\sqrt{3}}\~\approx\~0.9069\~.
\label{youhardwall01}
\eeq
\label{hardwalllemma01}
\end{lemma}

\noi
{\sc Sketched proof of Lemma~\ref{hardwalllemma01}}:~~Eq.\ \e{eyehardwall01} 
below follows directly from Lebesgue Majorant Theorem (LMT) using the second 
integral expression in \eq{eye01}. 
\beq
\lim_{E\to\infty} \frac{I(E)}{\sqrt{E}}\~\equi{\e{eye01}}\~
\lim_{E\to\infty} 2\int_{V\p_{0}}^{E} \sqrt{1-\frac{V}{E}}\~\ell^{\prime}(V)\~dV 
\=2\int_{V\p_{0}}^{\infty} \ell^{\prime}(V)\~dV
\=2\ell(\infty)\~<\~\infty\~.
\label{eyehardwall01}
\eeq 
[Note that it is easy to construct counterexamples to \eq{eyehardwall01} if
$V\p_{0}=-\infty$. Such counterexamples typically violate unitarity.] Similarly,
\beq
\lim_{E\to\infty}\sqrt{E} J(E)\~\equi{\e{jay01}}\~
\lim_{E\to\infty}\int_{V\p_{0}}^{E} 
\frac{\ell^{\prime}(V)\~dV}{\sqrt{1-\frac{V}{E}}}
\=\int_{V\p_{0}}^{\infty} \ell^{\prime}(V)\~dV\=\ell(\infty)\~,
\label{jayhardwall01}
\eeq 

\beq
\lim_{E\to\infty}\sqrt{E} K(E)\~\equi{\e{kay01}}\~
\lim_{E\to\infty}\int_{V\p_{0}}^{E} 
\frac{\ell(V)^{2}\ell^{\prime}(V)\~dV}{\sqrt{1-\frac{V}{E}}}
\=\int_{V\p_{0}}^{\infty} \ell(V)^{2}\ell^{\prime}(V)\~dV
\=\frac{\ell(\infty)^{3}}{3}\~,
\label{kayhardwall01}
\eeq
The \eqs{jayhardwall01}{kayhardwall01} do not follow directly from LMT per se, 
but possible (likely unphysical) counterexamples are beyond the scope of 
this article. Eq.\ \e{youhardwall01} is now a consequence of eqs.\ \e{you01}, 
\e{eyehardwall01}, \es{jayhardwall01}{kayhardwall01}.
\proofbox

\section{Bounded Potentials}
\label{secboundedpot}

\begin{lemma}[Bounded Potentials]
If the potential is bounded $-\infty<V\p_{0}\leq\Phi(x)\leq E\q_{0}<\infty$, 
then
\beq
U\~\gtrsim\~\frac{\pi}{2\sqrt{3}}
\~\approx\~0.9069\quad\for\quad N\~\gg\~1\~.
\label{youboundedpot01}
\eeq
\label{boundedpotlemma01}
\end{lemma}

\noi
{\sc Sketched proof of Lemma~\ref{boundedpotlemma01}}:~~Recall that we are 
still making the assumptions from Section~\ref{secassump} for simplicity. 
Bounded potentials are best analyzed directly in terms of the function
$0\leq x \mapsto \Phi(x)$ rather than the inverse function
$V\p_{0}\leq V \mapsto \ell(V)$ (up to factors of two). The independent
variable in the variation is the derivative $\Phi^{\prime}(x)\geq 0$ for
$x\geq 0$. The extremal profiles are finite square wells
\e{finitesquarewell01}, with the position $x=L/2$ of the (positive) kink as 
the only zeromode, which leads to the estimate \e{youboundedpot01}, \cf 
Appendix~\ref{secfinitesquarewell}.
\proofbox
 
\noi
{}Finally, The Lower Bound Theorem~\ref{thmlower} follows by use of 
Lemma~\ref{hardwalllemma01} and Lemma~\ref{boundedpotlemma01}, and the fact 
that there is no local minimum in the interior, \cf 
Sections~\ref{secextremalprofile}--\ref{secproofthmupper}.

\vspace{0.8cm}

\noi
{\sc Acknowledgement:}~K.B. would like to thank Tomas Tyc for fruitful
discussions. This article is inspired by question no.\ 88267 at the website
{\tt physics.stackexchange.com} asked by user Revo (user no.\ 4521).
The work of K.B.\ is supported by the Czech Science Foundation (GACR) under
the grant no.\ 14-02476S for Variations, Geometry and Physics.

\appendix

\section{Example: Positive Power Laws}
\label{secpospower}

\noi 
Let the potential be a positive power law
\beq
 \Phi(x)\=A \left(\frac{|x|}{L}\right)^{\frac{1}{\alpha}}+V\p_{0}\~,
\label{phipospower01}
\eeq
with $\alpha,A,L>0$ and $V,E\geq V_0$. Then the accessible length becomes
\beq
 \ell(V)\=2L \left(\frac{V-V\p_{0}}{A}\right)^{\alpha}\~.
\label{ellpospower01}
\eeq
The three integrals can be expressed in terms of the Euler Beta function 
\e{beta01}:
\beq
I\~\equi{\e{eye01}}\~
2L\frac{(E-V\p_{0})^{\alpha+\frac{1}{2}}}{A^{\alpha}} 
B(\alpha+1,\frac{1}{2})\~,
\label{eyepospower01}
\eeq
\beq 
J\~\equi{\e{jay01}}\~
2\alpha L\frac{(E-V\p_{0})^{\alpha-\frac{1}{2}}}{A^{\alpha}} 
B(\alpha,\frac{1}{2})\~,
\label{jaypospower01}
\eeq
\beq
K \~\equi{\e{kay01}}\~
8\alpha L^{3}\frac{(E-V\p_{0})^{3\alpha-\frac{1}{2}}}{A^{3\alpha}} 
B(3\alpha,\frac{1}{2})\~. 
\label{kaypospower01}
\eeq
Thus the \cp of uncertainties becomes
\beq
 U\~\stackrel{\e{you01}}{\approx}\~
\frac{\pi}{B(\alpha+1,\frac{1}{2})} 
\sqrt{\frac{B(3\alpha,\frac{1}{2})}{2(\alpha+\frac{1}{2}) 
B(\alpha,\frac{1}{2})}}\quad\text{for}\quad N\~\gg\~1.
\label{youpospower01}
\eeq
The relevant poles in the Euler Gamma function \e{beta01} are
\beq
 B(\alpha, \frac{1}{2})
\~\sim\~\frac{1}{\alpha}\quad\for\quad \alpha\~\to\~0\~,
\label{betanull}
\eeq
and (via the Stirling formula)
\beq
 B(\alpha, \frac{1}{2})\~\sim\~\sqrt{\frac{\pi}{\alpha}}
\quad\for\quad \alpha~\to~\infty\~. 
\label{betainfty}
\eeq

\noi
Remarks:
\begin{enumerate}

\item 
Positive power laws \e{youpospower01} respect the upper and lower bounds of
the main theorems from Section~\ref{secthm}.

\item
The infinite square well \e{inftywell01} corresponds to $\alpha=0$ with 
$U=\frac{\pi}{2\sqrt{3}}\approx 0.9069$.

\item
The harmonic oscillator \e{sho01} corresponds to 
$\alpha=\frac{1}{2}$ with $U=1$.

\item
The shallow potential corresponds to $\alpha=\infty$ with 
$U=\sqrt{\frac{\pi}{2\sqrt{3}}}\approx 0.9523$.
\end{enumerate}

\section{Example: Negative Power Laws}
\label{secnegpower}

\noi
Let the potential be a negative (attractive) power law
\beq
\Phi(x)\= E\q_{0}-A \left(\frac{L}{|x|}\right)^{\frac{1}{\alpha}}\~,
\label{phinegpower01}
\eeq
with $\alpha>\frac{1}{2}$, $A, L>0$, $V, E<E\q_{0}$, and $V\p_{0}=-\infty$. 
(One may show that the energy spectrum corresponding to 
$0<\alpha<\frac{1}{2}$ is unbounded from below, i.e. the system has no 
ground state. Hence we only consider $\alpha>\frac{1}{2}$.) The accessible 
length becomes
\beq
\ell(V)\=2L \left(\frac{A}{|E\q_{0}-V|}\right)^{\alpha}\~, 
\qquad V\~ <\~ E\q_{0} \~.
\label{ellnegpower01}
\eeq
The three integrals can again be expressed in terms of the Euler Beta 
function \e{beta01}:
\beq
I\~\equi{\e{eye01}}\~
2L\frac{A^{\alpha}}{|E\q_{0}-E|^{\alpha-\frac{1}{2}}} 
B(\alpha-\frac{1}{2},\frac{1}{2})\~,
\label{eyenegpower01}
\eeq
\beq
J\~\equi{\e{jay01}}\~
2\alpha L\frac{A^{\alpha}}{|E\q_{0}-E|^{\alpha+\frac{1}{2}}} 
B(\alpha+\frac{1}{2},\frac{1}{2})\~,
\label{jaynegpower01}
\eeq
\beq
K\~\equi{\e{kay01}}\~
8\alpha L^3\frac{A^{3\alpha}}{|E\q_{0}-E|^{3\alpha+\frac{1}{2}}} 
B(3\alpha+\frac{1}{2},\frac{1}{2})\~.
\label{kaynegpower01}
\eeq
Note that $N \propto I \to \infty$ for $|E\q_{0}-E|\to 0$, so that there are 
infinitely many bound states for negative power laws \e{phinegpower01}. 
(On top of that, there is a continuum of non-normalizable states $E>E\q_{0}$ 
which we are not interested in here.)

\noi
Thus the \cp of uncertainties becomes
\beq
U \~\stackrel{\e{you01}}{\approx}\~
\frac{\pi}{B(\alpha-\frac{1}{2},\frac{1}{2})}
\sqrt{\frac{B(3\alpha+\frac{1}{2},\frac{1}{2})}{2(\alpha-\frac{1}{2}) 
B(\alpha+\frac{1}{2},\frac{1}{2})}} \quad\for\quad N\~\gg\~1\~.
\label{younegpower01}
\eeq

\noi
Remarks:
\begin{enumerate}
\item
Negative power laws \e{younegpower01} respect the upper but not the lower 
bounds of the main theorems from Section~\ref{secthm}. (The lower bound does 
not apply since $V\p_{0}=-\infty$.)
 
\item
The shallow potential corresponds to $\alpha=\infty$ with 
$U=\sqrt{\frac{\pi}{2\sqrt{3}}}\approx 0.9523$, as we found previously.
\item
The inverse square potential corresponds to $\alpha=\frac{1}{2}$ with $U=0$. 
This is the threshold to quantum mechanically unstable Hamiltonians with 
spectrum unbounded from below. By going close to $\alpha=\frac{1}{2}$, it 
is possible to hide as many bound states (as we would like) down the throat, 
and compress them down to the theoretical limit given by the Heisenberg 
uncertainty principle (HUP).
\end{enumerate}

\section{Example: Finite Square Well}
\label{secfinitesquarewell}

\noi
The finite square well is
\beq
\Phi(x)\= V\p_{0} + ( E\q_{0}-V\p_{0}) \~\theta(L- 2|x|)
\=\left\{\begin{array}{lcl} V\p_{0}&\for& |x| \~<\~ \frac{L}{2}\~, \cr\cr 
E\p_{0}&\for& |x| \~>\~ \frac{L}{2}\~,\end{array}\right.
\label{finitesquarewell01}
\eeq
where  $V\p_{0}< E\q_{0}$ and $L>0$. The accessible length becomes
\beq
\ell(V)\~\equi{\e{conninvbranch01}}\~
L\~\theta(V-V\p_{0}) + \infty\~\theta(V-E\q_{0})
\=\left\{\begin{array}{lcl} 0 &\for& V \~<\~  V\p_{0}\~, \cr 
L &\for& V\p_{0} \~<\~  V \~<\~ E\p_{0}\~, \cr 
\infty &\for&   V \~>\~ E\p_{0}\~,
\end{array}\right.
\eeq
where we adopt the convention that $\infty\cdot 0=0$. The three integrals 
becomes
\beq 
 I(E)\~\equi{\e{eye01}}\~
4\int_{0}^{\Phi^{-1}(E)}\! \sqrt{E-\Phi(x)} \~dx
\~\equi{\e{finitesquarewell01}}\~
2L\sqrt{E-V\p_{0}} \~, \qquad V\p_{0}\~\leq\~ E\~ <\~ E\q_{0}\~, 
\label{eye02}
\eeq
\beq
 J(E)\~\equi{\e{jay01}}\~ 
2\int_{0}^{\Phi^{-1}(E)}\! \frac{dx}{\sqrt{E-\Phi(x)}}
\~\equi{\e{finitesquarewell01}}\~
\frac{L}{\sqrt{E-V\p_{0}}}\~,\qquad V\p_{0}\~\leq\~ E\~ <\~ E\q_{0}\~, 
\label{jay02} 
\eeq
\beq
 K(E)\~\equi{\e{kay01}}\~ 8\int_{0}^{\Phi^{-1}(E)}\!
\frac{x^{2}\~dx}{\sqrt{E-\Phi(x)}}
\~\equi{\e{finitesquarewell01}}\~
\frac{L^{3}}{3\sqrt{E-V\p_{0}}}\~, \qquad V\p_{0}\~\leq\~ E\~ <\~ E\q_{0}\~. 
\label{kay02} 
\eeq
Thus the \cp of uncertainties becomes
\beq
 U\~\stackrel{\e{you01}}{\approx}\~\frac{\pi}{2\sqrt{3}}\~\approx\~0.9069\~, 
\qquad V\p_{0}\~\leq\~ E\~ <\~ E\q_{0}\~.
\label{you02}
\eeq
Semiclassically, the \cp \e{you02} of uncertainties for the finite square
well agrees (not surprisingly) with the infinite square well \e{inftywell01}.

\section{Example: A Two-Stage Infinite Well}
\label{sectwostage}

\noi
The accessible length is
\beq
\ell(V)\= \sum_{i=0}^{1}L\q_{i} \~\theta(V-V\p_{i})
\=\left\{\begin{array}{lcl} 0 &\for& V \~<\~  V\p_{0}\~, \cr 
L\q_{0} &\for& V\p_{0} \~<\~  V \~<\~ V\p_{1}\~, \cr 
L\q_{0}+L\q_{1} &\for&   V \~>\~ V\p_{1}\~,
\end{array}\right.
\qquad L\q_{0},L\q_{1}\~\geq\~0\~, \qquad 
V\p_{0} \~\leq\~ V\p_{1}\~,
\label{elltwostage01} 
\eeq
where $\theta$ denotes the Heaviside step function. Eq. \e{elltwostage01}
corresponds to a two-stage infinite well potential
\beq
\Phi(x)\=V\p_{0} +(V\p_{1}-V\p_{0})\~\theta(2|x|-L\p_{0})
+\infty\~\theta(2|x|-L\p_{0}-L\p_{1})
\=\left\{\begin{array}{lcl} V\p_{0}&\for& |x| \~<\~ \frac{L\q_{0}}{2}\~, \cr 
V\p_{1}&\for& \frac{L\q_{0}}{2}\~<\~ |x| \~<\~ \frac{L\q_{0}+L\q_{1}}{2}\~, \cr
\infty&\for& |x| \~>\~ \frac{L\q_{0}+L\q_{1}}{2}\~.\end{array}\right. 
\eeq
The three integrals become 
\beq I(E)\~\equi{\e{eye01}}\~2\sum_{i=0}^{1}L\q_{i} \sqrt{E-V\p_{i}}
\~\theta(E-V\p_{i})\~, \label{eyetwostage01}
\eeq
\beq J(E)\~\equi{\e{jay01}}\~\sum_{i=0}^{1}L\q_{i}
\frac{\theta(E-V\p_{i})}{\sqrt{E-V\p_{i}}}\~,  \label{jaytwostage01}
\eeq
\beq K(E)\~\equi{\e{kay01}}\~\frac{L_{0}^{3}}{3}
\frac{\theta(E-V\p_{0})}{\sqrt{E-V\p_{0}}}
+\frac{(L\q_{0}+L\q_{1})^3-L_{0}^{3}}{3}
\frac{\theta(E-V\p_{1})}{\sqrt{E-V\p_{1}}}\~.\label{kaytwostage01}
\eeq
{}For fixed energy level $E>V\p_{1}$ and running $V\p_{0}\to -\infty$, 
the three pertinent integrals become 
\beq
\lim_{V\p_{0}\to-\infty} \frac{I}{\sqrt{-V\p_{0}}}
\~\equi{\e{eyetwostage01}}\~2L\q_{0}\~,
\label{eyetwostage02}
\eeq 

\beq
\lim_{V\p_{0}\to-\infty} J\~\equi{\e{jaytwostage01}}\~
\frac{L\q_{1}}{\sqrt{E-V\p_{1}}} \~,\qquad E\~>\~V\p_{1}\~,
\label{jaytwostage02}
\eeq

\beq
\lim_{V\p_{0}\to-\infty} K\~\equi{\e{kaytwostage01}}\~
\frac{(L\q_{0}+L\q_{1})^3-L_{0}^{3}}{3\sqrt{E-V\p_{1}}}\~,
\qquad E\~>\~V\p_{1}\~,
\label{kaytwostage02}
\eeq
The \cp \e{you01} of uncertainties $U\to 0$ vanishes in that limit
\beq
\lim_{V\p_{0}\to-\infty} U\~\equi{\e{you01}}\~0\~,
\qquad E\~>\~V\p_{1}\~,\qquad L\q_{1}\~>\~0\~.
\label{youtwostage02}
\eeq
Eq.\ \e{youtwostage02} shows that there is in general no non-zero lower bound
for the \cp \e{you01} of uncertainties $U$ for finite energy $E$ even if
the potential is bounded from below $\Phi(x)\geq V\p_{0}> -\infty$.

\section{Example: Logarithmic Potentials}
\label{secshallowpot}

\noi
Let the accessible length be of the form
\beq
\ell(V)\= P(V-V\p_{0})e^{\alpha (V-V\p_{0})}
\=P\left(\frac{d}{d\alpha}\right) e^{\alpha (V-V\p_{0})}\~, 
\qquad
\ell^{\prime}(V)\=P\left(\frac{d}{d\alpha}\right)\alpha e^{\alpha (V-V\p_{0})}\~,
\label{ellshallowpot01}
\eeq
where $\alpha>0$ is a positive constant and $P(z)=\sum_{k=0}^{m} a\q_{k} z^{k}$ 
is a polynomial with root $z=0$ (so that $\ell(V\p_{0})=0$). Let the energy 
level $E>V\p_{0}$ be arbitrary but fixed. We are interested in the shallow 
potential limit $\alpha\to \infty$. Concretely, let us assume that
\beq
\alpha\~\gg\~\frac{1}{E-V\p_{0}}\~.\label{bigalphashallowpot01}
\eeq
The three integrals then become Gaussian 
\bea
I&\equi{\e{eye01}+\e{ellshallowpot01}}&P\left(\frac{d}{d\alpha}\right)
\int_{V\p_{0}}^{E} \frac{e^{\alpha  (V-V\p_{0})}\~dV}{\sqrt{E-V}}
\~\stackrel{y=\sqrt{E-V}}{=}\~
2P\left(\frac{d}{d\alpha}\right)e^{\alpha  (E-V\p_{0})} 
\int_{0}^{\sqrt{E-V\p_{0}}} \!e^{-\alpha y^2}\~dy \cr
&\stackrel{\e{bigalphashallowpot01}}{\approx}& 
P\left(\frac{d}{d\alpha}\right)e^{\alpha (E-V\p_{0})}\sqrt{\frac{\pi}{\alpha}}
\~\stackrel{\e{bigalphashallowpot01}}{\approx}\~
\ell(E)\sqrt{\frac{\pi}{\alpha}}\~. 
\label{eyeshallowpot01}
\eea
Similarly,
\beq
J\~\stackrel{\e{jay01}+\e{ellshallowpot01}+\e{bigalphashallowpot01}}{\approx}\~
\ell(E)\sqrt{\pi \alpha} \~,
\label{jayshallowpot01}
\eeq
and
\beq
K\~\stackrel{\e{kay01}+\e{ellshallowpot01}+\e{bigalphashallowpot01}}{\approx}\~
\ell(E)^{3}\sqrt{\frac{\pi \alpha}{3}}\~. 
\label{kayshallowpot01}
\eeq
Note that $N \propto I \to \infty$ for $E\to\infty$, so that such logarithmic 
potentials \e{ellshallowpot01} have infinitely many bound states. The \cp 
\e{you01} of uncertainties becomes
\beq
U\~\stackrel{\e{you01}+\e{ellshallowpot01}+\e{bigalphashallowpot01}}{\approx}\~
\sqrt{\frac{\pi}{2\sqrt{3}}}~\approx~0.9523\quad\for\quad N\~\gg\~1\~. 
\label{youshallowpot01}
\eeq
Note that the \cp \e{youshallowpot01} of uncertainties has universal 
features in the sense that it doesn't depend on the parameters $E$, $V\p_{0}$, 
$\alpha$ (as long as \eq{bigalphashallowpot01} is satisfied), nor the 
polynomial $P$. 
The value \e{youshallowpot01} sits right in the middle of the double 
inequality.

\end{document}